\documentstyle[11pt,epsfig]{article}
\def\be{\begin{eqnarray}}
\def\ee{\end{eqnarray}}

\def\roughly#1{\mathrel{\raise.3ex\hbox{$#1$\kern-.75em%
\lower1ex\hbox{$\sim$}}}}

\begin{document}
\renewcommand{\thefootnote}{\arabic{footnote}}
\setcounter{footnote}{0}
\vskip 0.4cm


\hfill{\parbox{4cm}{KIAS-P99046\\HUTP-99/A036\\ \today}}

\vskip 1cm

\begin{center}
{\LARGE\bf Qualitons at High Density}
\date{\today}

\vskip 1cm

{\large
Deog Ki Hong$^a$\footnote{E-mail: dkhong@hyowon.cc.pusan.ac.kr},
Mannque Rho$^{b,c}$\footnote{E-mail: rho@spht.saclay.cea.fr} and
Ismail  Zahed$^{b,d}$\footnote{E-mail: zahed@nuclear.physics.sunysb.edu}
}
\end{center}

\vskip 0.5cm

\begin{center}

$^a$
{\it Lyman Laboratory of Physics,
Harvard University, Cambridge, MA 02138\\
and Department of Physics, Pusan National University,
Pusan 609-735, Korea}

$^b$
{\it School of Physics, Korea Institute for Advanced Study,
Seoul 130-012, Korea}

$^c$
{\it Service de Physique Th\'eorique, CE Saclay,
91191 Gif-sur-Yvette, France}

$^d$
{\it Department of Physics and Astronomy,
SUNY-Stony-Brook, NY 11794}

\end{center}

\vskip 0.5cm

\begin{abstract}
In the color-flavor-locked (CFL) phase of the QCD superconductor
we show that baryons behave as qualitons (called ``superqualitons")
with quantum numbers $B=(1\,\, {\rm mod}\,\, 2)/3$, $S=1/2$
and $Y=B$. An intriguing possibility implied by this identification
is that light baryonic modes in the form of superqualitons could be excited
below the (color) superconducting
gap in the CFL phase, a novel phenomenon foreign to normal
superconductors.
\end{abstract}

\newpage

\renewcommand{\thefootnote}{\#\arabic{footnote}}

\setcounter{footnote}{0}
{\bf 1.\,\,}
Dense hadronic matter poses still a challenging problem to theoretical
physicists despite years of vested efforts. With the advent of
dedicated fixed target accelerators (SIS) and heavy-ion colliders
(SPS, RHIC, LHC) the theoretical issues have become more imperative
as experiments may shed new
light and spur new interest in the problem.

At subnuclear matter densities, hadronic matter behaves as a free gas
of nucleons and a good understanding of this phase can be reached by
using virial type analyses~\cite{VIRIAL}. A liquid-gas or Ising
transition
has been proposed \cite{ISING} and recently observed~\cite{MSU}.
At a few times nuclear matter
densities, phenomenological analyses guided by nuclear structure data
and
constrained by neutron star observations have revealed a variety of
phenomena
including the possibility of a kaon-condensed phase that would be at
the origin of proton-rich stars~\cite{KAON}.
At densities larger than nuclear matter densities it has been recently
suggested that a variety of diquark-rich and superconducting phases
may take place with potentially large
gaps~\cite{NEWSUP,SON,pr98,HONG,hmsw99}.
Present neutron star
cooling scenarios however appear to disfavor large gaps at moderate densities,
although this conclusion is by no means definitive~\cite{PRAKASH}.

At asymptotic nuclear matter densities, quarks interact
weakly~\cite{perry} due to asymptotic freedom~\cite{gwp}.
The ensuing metallic phase is supposedly screened
except for possible infrared problems. It was originally suggested by
Bailin and Love~\cite{LOVE} and subsequently restressed
by others~\cite{NEWSUP} that in the
asymptotic regime quarks pair at the Fermi surface resulting in small
energy gaps. The perturbative arguments were recently revisited in
light of the fact that the magnetic modes undergo Landau
damping instead of screening, with the possibility of
large gaps~\cite{SON,pr98,HONG,hmsw99}. Here we just note
that perturbation theory may still fail at asymptotic densities due to
the nearness of the Gribov horizon in gauge theories~\cite{ZWANZIGER}.

Assuming that the superconducting phase is favored at asymptotic
densities, it was recently observed that for three massless flavors,
quarks may pair into diquarks with color and flavor locked
forming the CFL (color-flavor-locked) phase~\cite{arw99}.
It was further suggested that the CFL
phase exhibits a spectrum with matching quantum numbers to that of
the zero density phase~\cite{sw99}, a situation reminiscent of the
strongly coupled standard model~\cite{af81} and more intriguingly of
the Cheshire-Cat phenomenon in
low-energy hadronic structure (see \cite{RHO} and references therein).
The CFL phase appears to be energetically favored by effective theory
calculations~\cite{HONG}.

In this letter we show that in the superconducting CFL phase,
baryons behave as qualitons (called in short ``superqualitons"),
realizing
Kaplan's scenario in the high density
phase~\cite{KAPLAN}\footnote{We are
suggesting that the notion of quark soliton or qualiton
is more appropriate
at high density than at zero density at which the qualiton picture has
met with little success~\cite{karliner}}. They
carry quantum numbers $B=Y=(1\,\, {\rm mod}\,\, 2)/3$ and $S=1/2$, and
may even appear below the superconducting gap, a phenomenon unseen
in normal superconductors.
In section 2, we will give
arguments for the effective action in the CFL phase. In section 3,
we show that superqualitons are topologically stable and
identify their quantum numbers. In section 4, we
give two complementary
estimates for their mass and size. In section 5, we discuss
their quantized spectrum. Our conclusions are in section 6.

\vskip 1cm

{\bf 2.}
In the CFL superconducting phase, quarks with opposite Fermi momenta
pair with color and flavor locked. Model calculations~\cite{sw99}
and effective theory calculations~\cite{HONG} have shown that
this phase is favored energetically when the quark masses are
light for a reasonable choice of parameters.
The pertinent condensate will be taken in the
$(\overline{3},\overline{3})$ color-flavor representation
as
\be
\Big< q_{L\alpha}^{ia}q_{L\beta}^{jb} \Big>
=-\Big<q_{R\alpha}^{ia}q_{R\beta}^{jb}\Big>
=\kappa \,\,\epsilon^{ij}\epsilon^{abI}\epsilon_{\alpha\beta I}
\label{1}
\ee
where $\kappa$ is some constant,
$i,j$ are $SL(2,C)$ indices, $a,b$ are color indices,
and $\alpha,\beta$ are flavor indices. By allowing for an 
arbitrary relative phase between the left- and right-condensate,
parity could be spontaneously broken at high densities~\cite{pr98}.
For finite $\kappa$, both global color $SU(3)_c$
\footnote{The breaking of global color is of course a misdemeanor
that we will not address here. See \cite{arw99} and Langfeld and Rho
in \cite{NEWSUP} for discussions on
this point.} and flavor $SU(3)_{f,L,R}$ symmetries are broken.
The flavor-color locking in (\ref{1}) implies spontaneous breaking
through the color-flavor diagonal. Specifically:
$SU(3)_c\times\left( SU(3)_L\times SU(3)_R\right)\rightarrow
SU(3)_{c+L+R}$, with the emergence of 8 pseudoscalar
Nambu-Goldstone bosons together with 8 scalar Nambu-Goldstone
bosons eaten up by gluons through the Higgs mechanism.
There is an extra Nambu-Goldstone boson associated with
$U(1)_B\rightarrow Z_2$ with no relevance to our discussion.

To describe the low-energy dynamics of the color-flavor
locking phase, we introduce a field $U_L(x)$ which maps
space-time to the coset space,
$M_L=SU(3)_c\times SU(3)_L/SU(3)_{c+L}$.
Specifically,
the left-handed field is
\begin{equation}
{U_L}_{a\alpha}(x)\equiv\lim_{y\to x}{\left|x-y\right|^{\gamma_m}
\over\kappa}\epsilon^{ij}\epsilon_{abc}\epsilon_{\alpha\beta\gamma}
q^{b\beta}_{Li}(-\vec v_F,x)q^{c\gamma}_{Lj}(\vec v_F,y),
\end{equation}
where $\gamma_m$ is the anomalous dimension of the diquark field
of order $\alpha_s$
and $q(\vec v_F,x)$ denotes the quark field with momentum
close to a Fermi momentum $\mu\vec v_F$~\cite{HONG}.
The pairing involves quarks near the opposite edge
of the Fermi surface. Similarly, we introduce a right-handed
field $U_R(x)$, also a map from space-time to
$M_R=SU(3)_c\times SU(3)_R/SU(3)_{c+R}$,
to describe the excitations of the right-handed
diquark condensate.
Under an $SU(3)_c\times SU(3)_L\times SU(3)_R$
transformation by unitary
matrices $(g_c,g_L,g_R)$, $U_L$ transforms as $U_L\mapsto
g_c^*U_Lg_L^{\dagger}$ and $U_R$ transforms as
$U_L\mapsto g_c^*U_Rg_R^{\dagger}$.
In the ground state of the CFL superconductor,
$U_L$ and $U_R$ take the same constant value.
QCD symmetries imply
\begin{equation}
\left<{U_L}_{a\alpha}\right>=-\left<{U_R}_{a\alpha}\right>=\kappa\,
\delta_{a\alpha}.
\end{equation}
The Nambu-Goldstone bosons are the low-lying excitations of the
condensate, given as unitary matrices $U_L(x)=g_c^T(x)g_L(x)$
and $U_R(x)=g_c^T(x)g_R(x)$. 
For the present decomposition, we note the extra invariance under
the (hidden) local transformation $g_{c+L+R}(x)$ within 
the diagonal $SU(3)_{c+L+R}$ through
\be
g_c^T (x) \rightarrow g_c^T (x) g_{c+L+R}^{\dagger} (x)
\qquad\qquad
g_{L,R} (x)\rightarrow g_{c+L+R}(x) g_{L,R} (x) 
\label{local}
\ee
Hence, the spontaneous breaking of 
$SU(3)_c\times \left(SU(3)_L\times SU(3)_R\right)\rightarrow SU(3)_{c+L+R}$
can be realized non-linearly through the use of $U_{L,R}(x)$ or 
linearly through the use of $g_{c,L,R}(x)$ with the addition of an octet 
vector gauge field transforming inhomogeneously under local $g_{c+L+R}(x)$. 
This is the hidden local symmetry approach~\cite{KOICHI}, in which the 
ensuing vector gauge field is composite and Higgsed, which is to be
contrasted with a recent suggestion~\cite{WETTERICH}. 
Clearly the composites carry color-flavor
in the unbroken subgroup, with a mass of the order of the 
superconducting gap. Under some general 
conditions (e.g. vector dominance), the linear and nonlinear representations 
are the same~\cite{ULF}.  Some of these points will be further
discussed elsewhere.

The current associated with $\det U_L$ ($\det U_R$) is the
left (right)-handed $U(1)_L$ ($U(1)_R$)
baryon number current. Because of the $U(1)$ axial anomaly,
the field $\det U_L/ \det U_R$ is massive due to instantons
\footnote{The axial anomaly persists even in perturbation theory,
and does not fully require the presence of instantons. The instantons
of course persist even at asymptotic densities, albeit in a screened
form.}. We decouple the massive field from the low energy effective
action by imposing $\det U_L/ \det U_R=1$.
The Nambu-Goldstone boson
associated with spontaneously broken $U(1)_B$ symmetry is described
by $\det U_L\cdot\det U_R$ and responsible for baryon superfluidity.
But since it is not directly relevant for our problem, we further choose
$\det U_L\cdot\det U_R=1$ to isolate the Nambu-Goldstone bosons
resulting from the spontaneous breaking of chiral symmetry, from
the massive ones eaten up by the gluons. We now parameterize
the unitary matrices $U_{L,R}$ as
\begin{equation}
U_L(x)=\exp\left(2i \Pi_L^AT^A/F\right),
\quad U_R(x)=\exp\left(2i\Pi_R^AT^A/F\right),
\end{equation}
where $T^A$ are $SU(3)$ generators, normalized as
${\rm Tr}~T^AT^B=\delta^{AB}/2$. The Nambu-Goldstone bosons
$\Pi^A$ transform nonlinearly under
$SU(3)_c\times SU(3)_L\times SU(3)_R$ but
linearly under the unbroken symmetry group $SU(3)_{c+L+R}$.

The effective Lagrangian for the CFL phase
at asymptotic densities follows by integrating out the `hard' quark
modes at the edge of the Fermi surface as suggested by Polchinski
\cite{POLCHINSKY} and Weinberg
\cite{SUPERWEINBERG}. This program
has been carried out partly
by one of us~\cite{HONG}. The effective
Lagrangian for $U_{L,R}$ is a standard non-linear sigma model in d=4
dimensions and should include the interaction of Nambu-Goldstone
bosons with colored but ``screened" gluons $G$
\footnote{The magnetic gluons are Landau-damped in perturbation
theory, but this may not be valid at asymptotic
densities as suggested in \cite{ZWANZIGER},
if $m_*\sim (m_E^2\Delta)^{1/3}$.}.
The $SU(3)_c$ current of
Nambu-Goldstone bosons in the CFL phase
consists of two pieces; one is from the Noether color-current
and the other one is from the WZW term,  given as
\begin{equation}
{J_{cL}}^{A\mu}={i\over 2}F^2{\rm Tr}~U_L^{-1}T^A\partial^{\mu}U_L+
{1\over 24\pi^2}\epsilon^{\mu\nu\rho\sigma}
{\rm Tr}~T^AU_L^{-1}\partial_{\nu}U_LU_L^{-1}\partial_{\rho}U_L
U_L^{-1}\partial_{\sigma}U_L,
\end{equation}
where the first term is the Noether current and the second one is
from the WZW term.
Expanding in powers of derivative, the effective Lagrangian for
the (colored) Nambu-Goldstone bosons is then
\begin{eqnarray}
{\cal L} =&+&
\frac {F_T^2}4 \,\,{\rm Tr} (\partial_0 U_L\partial_0 U_L^{\dagger})
-\frac {F_S^2}4 \,\,{\rm Tr} (\partial_i U_L\partial_i U_L^{\dagger})
+g_sG\cdot J_{cL}+n_L{\cal L}_{WZW}
\nonumber\\
&+ &  (L\rightarrow R)+
{\cal O} \left( \frac {\partial^4}{(4\pi\Delta)^4}\right).
\label{2x}
\end{eqnarray}
In the presence of a chemical potential, only $O(3)$ is enforced in
space. In carrying the derivative expansion, we assume the superconducting
scale  $4\pi\Delta$ to be the large scale. (In effective theories the
emergence of $4\pi$ is purely a loop effect in d=4).
The Wess-Zumino-Witten~\cite{witten1,wz71} (WZW) term is
nonlocal in d=4. Its normalization will be discussed below. The
temporal and spatial decay constants $F_{T,S}$ are fixed by the
`hard' modes at the Fermi surface. Their exact values are determined
by the dynamics at the Fermi surface. For simplicity, we will assume
$F_S\sim F_T\sim F\sim\Delta$ from now on.

We note that (\ref{2x}) only accounts for the gluon interaction to
leading order in $g_s$. To order $g_s^2$ the gluon
mediated interaction yields $L,R$ mixing in a parity-invariant way. 
The breaking of chiral symmetry in the CFL phase via
composite operators, e.g. 
$\left<(\overline q q)^{2n}\right>\sim |\left<qq\right>|^{2n}$,
does not imply additional order parameters. No
additional Goldstone modes are needed as all expectation values
follow through simple Fierz rearrangements.

The effective Lagrangian (\ref{2x}) in the CFL phase
bears much in common with
Skyrme-type Lagrangians~\cite{skyrme,adkins83,RHO}, with the
screened color mediated interaction analogous to the exchange
of massive vector mesons. Indeed in the CFL phase the `screened'
gluons and the `Higgsed' gauge composites are the analogue of the
massive vector mesons in the low density phase~\cite{sw99}. 
In the CFL phase the WZW term
is needed to enforce the correct flavor anomaly structure. Its
structure is best described in 5-dimensional space as the
homotopy of the Nambu-Goldstone manifold $M$, $\Pi_5(M)=Z$.
Through the normalization $S_{WZW}=2m\pi$ with $m\in Z$,
its coefficient $n_L$ and $n_R$ must be an integer
to have a consistent
quantum theory.

\vskip 1cm
{\bf 3.\,\,}
Much like the effective Lagrangian for QCD at low density (giving rise
to skyrmions), the
low-energy effective Lagrangian in the CFL phase admits a
stable (static) soliton solution, with a
winding number given by the homotopy $\Pi_3(M)=Z$.
It is stable by the balance between the kinetic energy
(attractive force)
and the Coulomb energy (repulsive force)~\cite{hr90}.\footnote{There
can be additional higher derivative terms that may significantly
contribute to the stabilization of the qualiton which we shall not
address here. Differently to the qualiton in the vacuum, one would also
have to consider terms that are specific to the Fermi surface.}  Since
the soliton quantum numbers are determined by the
WZW term upon quantization, we need to
determine the coefficient $n$. For skyrmions,
Witten~\cite{witten1} has shown that $n_L$ and $n_R$ are
fixed by the underlying
flavor anomalies in QCD. Here we proceed to show that $n_L$ and $n_R$
are fixed by analogous color-flavor anomalies in the CFL phase.

First, consider the anomalous $SU(3)_L$ current,
the pertinent $U(1)$ subgroup of which is usually
at the origin of the pion decay into two photons
(i.e., ABJ triangle-anomaly).
Since under the $U(1)_{\rm em}$ electro-magnetic transformation
the quark transforms as $q\mapsto e^{i\epsilon Q}q$
with $Q={\rm diag}~(2/3,-1/3.-1/3)$, the $U(1)_{\rm
em}$ transformation of the CFL condensate can be
undone by a $U(1)_Y$ hyper-charge transformation of $SU(3)_c$ with
$Y=-Q$. Therefore, the unbroken
$U(1)_{\tilde Q}$ gauge boson (or modified photon) is a
linear combination of the original photon and gluon~\cite{WILCZEK}
\begin{equation}
{\tilde A}_{\mu}=A_{\mu}\cos\theta+G_{\mu}^Y\sin\theta
\end{equation}
where $\cos\theta=g_s/\sqrt{e^2+g_s^2}$ and $G_{\mu}^Y$ is the gluon
field for $U(1)_Y\subset SU(3)_c$. Since the quark now couples to
the modified photon with strength $\tilde
e=eg_s/\sqrt{e^2+g_s^2}$, the unit charge of $U(1)_{\tilde Q}$ is
$\tilde e$.

A $U(1)_{\tilde Q}$ transformation
$\exp(i\epsilon {\tilde Q})$ on $U$ is nothing but a simultaneous
transformation of $g_c=\exp(i\epsilon Y)$ and
$g=\exp\left(i\epsilon Q\right)$:
\begin{eqnarray}
U\mapsto e^{i\epsilon Q}Ue^{-i\epsilon Q},
\label{uone}
\end{eqnarray}
where we have used $Y=-Q$ for the left multiplication.
We see that the $U(1)_{\tilde Q}$ transformation Eq.~(\ref{uone})
is identical to the $U(1)_{\rm em}$ transformation of the
Nambu-Goldstone bosons associated
with chiral symmetry breaking $SU(3)_L\times SU(3)_R\mapsto
SU(3)_V$ in the QCD vacuum, except for the fact that the unit charge
is now $\tilde e=eg_s/\sqrt{e^2+g_s^2}$ instead of $e$.

Therefore, all the modified electric charges of the Nambu-Goldstone
bosons in the CFL phase
are integral multiple of $\tilde e$, the modified electric
charge of the electron. Note that all the quarks and gluons have
integer (modified) electric charges too,  since quarks and gluons
transform as Eq.~(\ref{uone}) under $U(1)_{\tilde Q}$~\cite{arw99}.
Specifically,
\begin{eqnarray}
G_{\mu}& \mapsto & e^{i\epsilon {\tilde Q}}G_{\mu}e^{-i\epsilon
{\tilde Q}}=e^{i\epsilon Y}G_{\mu}e^{-i\epsilon Y}
\\
q& \mapsto & e^{i\epsilon {\tilde Q}}qe^{i\epsilon
{\tilde Q}}=e^{i\epsilon Y}qe^{i\epsilon Q}
\end{eqnarray}

The $SU(3)_L$ anomaly is now given as
\begin{equation}
\partial^{\mu}{{J_L}_{\mu}^A}={{\tilde e}^2\over 32\pi^2}{\rm Tr}
\left(T^A{\tilde Q}^2\right)\epsilon_{\mu\nu\rho\sigma}{\tilde
F}^{\mu\nu}{\tilde F}^{\rho\sigma},
\label{anomaly}
\end{equation}
where $\tilde F^{\mu\nu}$ is the field strength tensor of the
modified photon associated with $U(1)_{\tilde Q}$.
Note that there is no color or flavor
factor in the coefficient because neither color nor flavor is
running around the triangle diagram responsible for the
$SU(3)_L$ anomaly. In the CFL phase color and flavor are locked,
and the modified pion and photon carry color. In the CFL phase,
\begin{equation}
{\tilde Q_{a\alpha}}=\left(\begin{array}{c c c}
                      0&-1&-1 \\
                      1&0&0 \\
                      1&0&0
                 \end{array}  \right),
\end{equation}
and only three quarks contribute to the anomaly in the
${J_L}_{\mu}^3$ current:
${\rm Tr}\left(T^3{\tilde Q}^2\right)=1/2\cdot\left(1^2+1^2\right)
+(-1/2)\left(-1\right)^2=1/2$.

To fix the coefficient $n$ of the WZW term from the $SU(3)_L$ anomaly,
we introduce the
modified electro-magnetic interaction by gauging the
$U(1)_{\tilde Q}$ symmetry in the effective action (\ref{2x})
through
\begin{equation}
\partial^{\mu} U\mapsto D^{\mu}U_{a\alpha}=\partial^{\mu}U_{a\alpha}
-i{\tilde e}{\tilde A}_{\mu}\left(Y_{ab}U_{b\alpha}+
U_{a\beta}Q_{\beta\alpha}\right).
\end{equation}
As shown by Witten~\cite{witten1}, the $U(1)_{\tilde Q}$
invariant WZW term then contains a term
\begin{equation}
S_{WZW}(U,{\tilde A}_{\mu})\ni {n_L\tilde e^2\over 64\pi^2 F}
\Pi^0\epsilon^{\mu\nu\alpha\beta}{\tilde F}_{\mu\nu}{\tilde
F}_{\alpha\beta},
\end{equation}
which agrees with (\ref{anomaly}) if $n_L=1$. Similarly, one can
also show that $n_R=1$.

On the other hand, the topologically conserved current for
the soliton in (\ref{2x}) is
\begin{equation}
J_L^{\mu}={1\over 24\pi^2}\epsilon^{\mu\nu\rho\sigma}
{\rm Tr}~U_L^{-1}\partial_{\nu}U_LU_L^{-1}\partial_{\rho}U_LU_L^{-1}
\partial_{\sigma}U_L.
\label{topc}
\end{equation}
Since the sigma model description of the $SU(3)_L$ quark current
${J^{A}_{L\mu}}=\bar q_LT^A\gamma_{\mu}q_L$ contains an anomalous
piece from the WZW term
\begin{equation}
{J^{A}_{L\mu}}\ni{1\over 24\pi^2}\epsilon^{\mu\nu\rho\sigma}
{\rm Tr}~T^A{U_L}^{-1}\partial_{\nu}{U_L}{U_L}^{-1}
\partial_{\rho}{U_L}{U_L}^{-1}\partial_{\sigma}{U_L}
\end{equation}
and the topological current Eq.~(\ref{topc})
corresponds to the anomalous piece of the $U(1)_L$ current
in the $\sigma$ model description~\cite{witten1,witten2,bal83},
we find the soliton of unit winding number has a quark number $1$
or baryon number $B=1/3$.
This is Kaplan's qualiton~\cite{KAPLAN} in
the CFL phase. We are referring to it as the superqualiton
to distinguish it from Kaplan's qualiton in the vacuum. Since in the
CFL phase baryon number is spontaneously broken $U_B(1)\rightarrow Z_2$,
it is clear that $B=1/3$ is the same as $B=1$,
hence $B=(1\,{\rm mod}\,2)/3$.
The fact that $n_L=1$ (or $n_R=1$)
implies that the superqualiton is a fermion.
The superqualiton carries positive parity, since the
CFL condensate is parity even.
Further quantum numbers along with the excited spectrum of the
superqualiton will be discussed below.

\vskip 1cm

{\bf 4.\,\,\,}
Some important questions regarding  superqualitons concern their mass
$M_S$ and their interactions. Indeed, if $M_S$ happens to be less than
$\Delta$ (where $\Delta$ is the superconductivity gap), then
superqualitons may exist inside the superconducting CFL gap with
ordinary baryon quantum numbers, a phenomenon unseen in conventional
superconductors. Attractive interactions among light superqualitons
may even trigger a total rearrangement inside the QCD superconductor.

A simple estimate of the mass follows from scale (virial-like)
arguments that are in general not specific to the detailed form of
(\ref{2x}).
Indeed, for stable static configurations a typical gradient in
(\ref{2x}) is $\partial\sim \Delta$. Since $F\sim \Delta$
and the size of the superqualiton $R_S\sim 2\pi/(4\pi\Delta)$,
then
\be
M_S\sim \left(\frac 12 F^2\Delta^2\right)\,\,
\left( \frac 43 \pi R_S^3\right)
\sim \frac 14\Delta
\label{estimate}
\ee
which is less than $\Delta/2$. Both the mass and the size are fixed
by $\Delta$. To create 2 superqualitons costs
less than to excite 2 bare quarks from the CFL phase. Such
superqualitons may either condense (crystallize) into a new
superconductor with a reduced gap, or exhibit a liquid or gas
phase with either normal or superconducting properties. In order
to address these issues, the specific dynamics of the many-superqualiton
interaction is required. Unfortunately we have presently no good
understanding of this aspect of the problem.

A more conservative estimate of the mass of the superqualiton
can be made by relying on the effective Lagrangian (\ref{2x}).
Following Skyrme~\cite{skyrme} and
others~\cite{RHO} (and references therein) we seek a static
configuration for the field $U_L$ in $SU(3)$ by embedding an
$SU(2)$ hedgehog in color-flavor in $SU(3)$, with
\begin{equation}
{U_L}_c(x)=e^{i\vec\tau\cdot \hat r\theta(r)},\quad U_R=0,
\quad G^Y_0=\omega(r),\quad {\rm all~~ other}~~G_{\mu}^A=0,
\label{profile}
\end{equation}
where $\tau$'s are Pauli matrices. The radial function $\theta(r)$
is monotonous and satisfies
\begin{equation}
\theta(0)=\pi,\quad\theta(\infty)=0
\label{bound}
\end{equation}
for a soliton of winding number one.
(Note that we can also look for a right-handed soliton by switching
off the $U_L$ field. The solution should be identical because
(\ref{2x}) is invariant under parity.) This configuration
has only non-vanishing color charge in the $Y$ direction
\begin{equation}
J_{0}^Y={\sin^2\theta~\theta^{\prime}\over 2\pi r^2}
\end{equation}
and all others are zero.
As shown in~\cite{hr90}, the energy of the configuration
(\ref{profile}) is given as
\begin{equation}
E[\omega,\theta]=\int 4\pi r^2{\rm d}r\left[
-{1\over2}{\omega^{\prime}}^2+F^2\left({\theta^{\prime}}^2
+2{\sin\theta^2\over r^2}\right)+{g_s\over 2\pi^2}
{\omega\over r^2}\sin^2\theta~\theta^{\prime}
\right].
\end{equation}
The total charge within a radius $r$ is
\begin{equation}
Q^Y(r)=g_s\int_0^r {\rm Tr}~YJ_0^Y(r^{\prime})4\pi
{r^{\prime}}^2{\rm d}r^{\prime}
=-g_s\left({\theta(r)-\sin\theta(r)\cos\theta(r)-\pi\over\pi}
\right).
\end{equation}
Using Gauss law with a screened charge density,
we can trade  $\omega$ in terms of $\theta(r)$,
\begin{equation}
\omega^{\prime}={Q^Y(r)\over 4\pi r^2}e^{-m_Er},
\end{equation}
where $m_E=\sqrt{6\alpha_s/\pi}\,\mu$ is the electric screening mass
for the gluons (note that the magnetic gluons are not needed
for the static configuration).
Hence, the energy functional simplifies to
\begin{equation}
E[\theta]=\int_0^{\infty}{\cal E}(r) \, {\rm d}r =
\int_0^{\infty}4\pi{\rm d}r\left[
F^2\left(r^2{\theta^{\prime}}^2+2\sin^2\theta\right)
+{\alpha_s\over2\pi}\left({\theta-\sin\theta\cos\theta-\pi
\over2r}\right)^2e^{-2m_Er}
\right],
\end{equation}
where $\alpha_s=g_s^2/(4\pi)$.
The squared size of the superqualiton is $R_S^2= \left<r^2\right>$
where the averaging is made using the (weight) density
${\cal E}(r)$.
The equation of motion for the superqualiton profile $\theta (r)$ is
\be
\left(r^2\theta'\right)' ={\rm sin}2\theta +
\frac {\alpha_s}{4\pi}\frac{e^{-2m_Er}}{(F\, r)^2}\sin^2\theta
\left(\pi -\theta+\frac 12 {\rm sin}2\theta \right)
\label{mot}
\ee
subject to the boundary conditions (\ref{bound}).

We have solved (\ref{mot})
numerically for several values of $m_E$ and $\alpha_s$.
Given a fixed $\alpha_s=1$, we find the soliton mass
$M_S=1.53,~1.06,~0.75\times 4\pi F$ and
the soliton radius $R_S=0.29,~0.20,~0.15~F^{-1}$
for $m_E/F=2, 5, 10$, respectively. 
For various values of $\alpha_s=1.0,~0.5,~0.2$ with
a given $m_E/F=2$, we get
$M_S=1.53,~1.21,~0.87\times 4\pi F$ and
$R_S=0.29,~0.24,~0.17~F^{-1}$. 
We see that both the mass and the size of the soliton decrease as
the repulsive Coulomb force decreases, showing that
the soliton gets smaller for the weaker coupling, since
the color-electric force, which balances the kinetic energy
of the soliton, is less repulsive~\cite{hr90}.
For the high density region, where the QCD coupling is small,
we expect the higher order derivative terms to become more important
in stabilizing the soliton, since the Coulomb forces become negligible
following screening (an alternative would be through the use of 
gauge composites). This trend is expected from the screened character 
of the interaction, and seems to differ from the one we suggested earlier 
by our simple scaling arguments. Which scenario (among the various alternatives)
is favored will depend on the content of the effective Lagrangian implied
by first principles of QCD which may not be reliably embodied in
(\ref{2x}).

\vskip 1cm
{\bf 5.\,\,}
To access the quantum numbers and the spectrum of the superqualiton,
we note as usual that for any static solution to the equations
of motion,
one can generate another solution by a rigid $SU(3)$ rotation,
\begin{equation}
U(x)\mapsto A U(x)A^{-1},\quad A\in SU(3).
\end{equation}
The matrix $A$ corresponds to the zero modes of the superqualiton.
Note that two $SU(3)$ matrices are equivalent if they differ by
a matrix $h\in U(1)\subset SU(3)$ that commutes with $SU(2)$
generated by $\vec \tau\otimes I$.
The Lagrangian for the zero modes is given by substituting
$U(\vec x,t)=A(t)U_c(\vec x)A(t)^{-1}$~\cite{bal85}. Hence,
\begin{equation}
L[A]=-M_S+{1\over2}I_{\alpha\beta}{\rm Tr}~T^{\alpha}A^{-1}
\dot{A}{\rm Tr}~T^{\beta}A^{-1}{\dot A}
-i{1\over 2}{\rm Tr}~YA^{-1}{\dot A},
\end{equation}
where $I_{\alpha\beta}$ is an invariant tensor on
${\cal M}=SU(3)/U(1)$ and the hypercharge $Y$ is
\begin{equation}
Y={1\over3}\pmatrix{1&0&0\cr
0&1&0\cr
0 & 0 &-2}.
\end{equation}
Under the transformation $A(t)\mapsto A(t)h(t)$ with
$h\in U(1)_Y$
\begin{equation}
L\mapsto L-{i\over2}{\rm Tr}~Yh^{-1}\dot h.
\end{equation}
Therefore, if we rotate adiabatically the soliton by $\theta$ in the
hypercharge space in $SU(3)$, $h=\exp(iY\theta)$,
for time $T\to\infty$, then the wave
function of the soliton changes by a phase in the semiclassical limit:
\begin{equation}
\psi(T)\sim e^{i\int {\rm d}tL}\psi(0)=e^{i\theta/3}\psi(0),
\label{wfn}
\end{equation}
where we neglected the irrelevant phase $-M_ST$ due to the rest mass
energy. The simplest and lowest energy configuration that satisfies
Eq.~(\ref{wfn}) is the fundamental representation of $SU(3)$.
Similarly, under a spatial (adiabatic) rotation by $\theta$
around the $z$ axis, $h=\exp(i\tau^3\theta)$,
the phase of the wave function changes by $\theta/2$.
Therefore, the ground state of the soliton is
a spin-half particle transforming under the fundamental representation
of both the flavor and the color group, which leads us to conclude that
the soliton is a massive left-handed (or right-handed) quark in the CFL
phase. 

It is interesting to note that the phase continuity between
high-density quark matter and low-density nuclear matter, conjectured
by Sch\"afer and Wilczek~\cite{sw99}, carries to
the Skyrme picture of baryons. One way to address the specifics of the
continuity besides a spectrum analysis, is via anomaly-matching.
This issue will be addressed elsewhere.

\vskip 1cm
{\bf 6.\,\,}
In the CFL phase of the QCD superconductors, baryons emerge
as qualitons (superqualitons).
We have shown that the quantum numbers of the superqualitons
are dictated by a WZW term suitably normalized by the
CFL triangle-anomaly. We have presented general arguments for
the effective Lagrangian in the CFL phase. Superqualitons can
be stabilized by screened colored electric interactions for
a variety of chemical potentials, exhibiting a rich
and colorful spectrum. Our estimate of the mass is on the
higher side of the superconducting gap, although simple
scaling arguments show it otherwise.
Our calculations do not exhaust all parameter space for
effective model calculations in QCD superconductors, and
point at the necessity of a first principle assessment
of the effective Lagrangian (\ref{2x}). One fascinating possibility
suggested by the superqualiton description of the hadron-quark continuity
is that light fermionic modes  with $M_s<\Delta$ get excited
within the CFL superconducting gap, a
phenomenon unseen in normal superconductors.
On general grounds, the interaction between
light superqualitons may cause a rearrangement
in the superconducting phase with
important implications for the bulk
properties such as the specific heat. The latter is
paramount to cooling curves of neutron stars.

\vskip 0.2in

{\bf Acknowledgments}

\vskip .35cm
We are grateful to Byung-Yoon Park,
Hanchul Kim and Claudio Rebbi for discussions and 
valuable help in the numerical analysis and to
Krishna Rajagopal, Martin Rocek and
Thomas Sch\"afer for comments on the manuscript. 
MR and IZ thank KIAS for their generous hospitality and support during the
completion of this work. 
DKH acknowledges the financial support of
the Korean Research Foundation program 1998-15-D00022.
The work of IZ was partially supported by the  US-DOE
grant DE-FG-88ER40388.

\vskip .25cm
{\it Note added}
\vskip .25cm
After posting our paper, \cite{GATTO} appeared in which some of the
present issues were discussed. We note that the linear interpretation 
of the symmetry breaking pattern adopted by the authors in \cite{GATTO}
differs from ours and that our nonlinear 
description {\it does} contain $L,R$ mixing to order $g_s^2$ contrary 
to their statement.

\eject











\end{document}